\documentstyle[twocolumn,prl,aps]{revtex}
\begin{document}
\draft

\title{Quantum Ferromagnetism and Phase Transitions in Double-Layer Quantum
Hall Systems}

\author{Kun~Yang$^1$, K.~Moon$^1$, L.~Zheng$^1$, A.H.~MacDonald$^1$,
S.M.~Girvin$^1$, D.~Yoshioka$^2$, and Shou-Cheng Zhang$^3$}
\address{$^1$Department of Physics, Indiana University, Bloomington,
IN~~47405\\
$^2$Institute of Physics, College of Arts and Sciences, University of
Tokyo, Komaba, Meguroku Tokyo 153, Japan\\
$^3$IBM Alamaden Research Center, San Jose, CA~~95120
and
Department of Physics, Stanford University, Palo Alto, CA~~94305}

\date{\today}

\maketitle
\begin{abstract}
Double layer quantum Hall systems have interesting properties associated with
interlayer correlations.  At $\nu =1/m$ where $m$ is an odd integer they
exhibit spontaneous symmetry breaking
equivalent to that of spin $1/2$ easy-plane ferromagnets,
with the layer degree of freedom playing the role of spin.
We explore the rich variety of quantum and finite temperature phase transitions
in these systems.  In particular, we show that a magnetic field oriented
parallel to the layers induces a highly collective commensurate-incommensurate
phase transition in the magnetic order.
\end{abstract}

\pacs{75.10.-b, 73.20.Dx, 64.60.Cn}

\narrowtext
Recent technological progress has allowed production of double-layer
quantum
Hall systems of extremely high mobility.
The separation $d$ of
the two 2D electron gases is so small ($d\sim 100$\AA ) as to be comparable
to
the spacing between electrons in the same layer
and quantum states with strong correlations
between the layers have been observed experimentally and discussed
theoretically
\cite{exp,theory,murphytobepub}.
Wen and Zee have pointed out that at Landau level
filling
factor
$\nu = 1/m$ and in the absence of interlayer tunneling, this system
exhibits a
spontaneously broken U(1) gauge symmetry\cite{wenandzee}.
($m$ is an odd integer.)
The corresponding Goldstone mode
is a
neutral density wave in which the densities in the two layers oscillate out
of
phase.
A finite temperature Kosterlitz-Thouless (KT) phase transition
is expected to be associated with this broken symmetry.

In this paper we focus for convenience
on the case $\nu =1$ and show that this system can be
viewed as an {\it easy-plane
quantum  itinerant ferromagnet}.
Following
Ref.~\cite{macd-platzman-boebinger} (but with a rotated coordinate system)
we
will use an `isospin' magnetic language in which isospin `up' (`down') refers
to an
electron in the `upper' (`lower') layer \cite{footnote}.  Using this
language
and building upon recent progress in understanding the case of single-layer
systems at $\nu =1$ with real spin \cite{leeandkane,sondhi} we explore the
consequences of
the mixing of charge and isospin degrees of freedom and discuss the rich
variety
of phase transitions controlled by temperature, layer separation,
tunneling between layers, layer charge
imbalance and magnetic field tilt angles.
In addition to the KT transition we
find
a `commensurate-incommensurate' phase transition as a function of
$B_\parallel$,
the component of the magnetic field in the plane.  Furthermore we
demonstrate
that the Meissner screening of the in-plane component of the
magnetic field ($B_\parallel$) predicted by Ezawa and Iwazaki
does
{\it not\/} occur.  A portion of this rich set of phenomena is captured in
the schematic
zero-temperature phase diagram illustrated in Fig.~\ref{fig1}.
The
present paper will be devoted to explication of the physical picture
underlying
this phase diagram.  Technical details of the microscopic calculations on
which
it is based will be presented elsewhere \cite{ustobepub}.

It is helpful to begin by discussing the limit of zero
temperature, zero tunneling amplitude between the layers and layer
spacing $d=0$.  We
work entirely in the lowest Landau level and take the unit of length to be
$l\equiv (\hbar c/eB)^{1/2}$.  Coulomb repulsion induces a strong exchange
interaction $E_{\rm x}\sim 60{\rm K}~\sqrt{B/{\rm Tesla}}$ which seeks to
ferromagnetically align all the isospins.  Because the kinetic energy has
been quenched by the Landau level degeneracy, the exact ground state is
simply a
single Slater determinant consisting of a filled Landau level with all
electrons fully polarized in the same isospin state.  Since in this special
limit the interactions do not depend on  which layer the particles are
in,
the system exhibits full $SU(2)$ invariance $[H,S^{\mu}]=0$ and the
direction
of
the polarization is arbitrary.
 The total isospin for $N$ electrons is $S=N/2$ and the
$2S +
1$ eigenstates of $S^z\equiv (N_\uparrow - N_\downarrow )/2$ are completely
degenerate.

The exact single-magnon excited states can also be
found \cite{magnon,macd-platzman-boebinger}.  Taking the
magnon vacuum to be the $S^z = -N/2$ ground state, we have
\begin{equation}
\vert {\bf k}\rangle = \overline{S_{\bf k}^+}~\vert 0\rangle
\label{eq70}
\end{equation}
where $S_{\bf k}^+$ is the Fourier transform of the isospin raising
operator
and the overbar indicates projection onto the lowest Landau level.
In the single-magnon subspace there is only one state for each
wavevector so that $| \bf k \rangle$ is an exact eigenstate.
The eigenvalue
$\epsilon_{\bf k}$ vanishes quadratically at small wavevectors and
saturates at
$E_x$ at large wavevectors.  The quadratic dispersion at small $k$ implies
an
effective isospin stiffness energy for variations in the unit-vector order
parameter field $m^\mu ({\bf r})\equiv (4\pi\ell^2)\langle S^\mu ({\bf
r})\rangle$
\begin{equation}
T=\frac{1}{2}~\rho_s~\int d^2r\> ({\bf \nabla}m^\mu )\cdot ({\bf
\nabla}m^\mu )
\label{eq100}
\end{equation}
with $\rho_s = e^2/(16\sqrt{2\pi}\epsilon\ell)$.
The microscopic origin of this stiffness is the loss of exchange energy
when the
spin orientation varies with position\cite{sondhi,ustobepub}.

These properties of the ground state and small wave vector excited states
are
similar to those of the 2D quantum ferromagnetic Heisenberg model.
However,
this system is most comparable to
itinerant electron ferromagnets and this has some especially interesting
consequences \cite{sondhi,ustobepub}.  In particular at $\nu = 1$ charged
excitations can only be created by distorting the isospin orientation and,
quite remarkably, the fermionic charge density is simply the Pontryagin
topological
density \cite{sondhi,ustobepub}
\begin{equation}
\delta\rho ({\bf r}) = \frac{1}{8\pi}~ \epsilon^{\mu\nu}~ {\bf m}\cdot
(\partial_\mu~
{\bf
m})\times (\partial_\nu~ {\bf m}).
\label{eq150}
\end{equation}
Thus even though the isospin spectrum is gapless, the isospin stiffness implies
a
finite charge gap, the lowest energy state carrying charge being a
`skyrmion'
spin texture \cite{sondhi,ustobepub} whose energy is one-half of the
Hartree-Fock charge excitation energy (in the SU(2) invariant case).

There are several ways in which the $SU(2)$ symmetry can be lifted.  The
most important consequence of finite
layer separation $d$ is the creation of
a local capacitive charging energy which for slowly varying charge density
($m^z$) is \cite{wenandzee,macd-platzman-boebinger}
\begin{equation}
U_z = \frac{e^2 d \, \eta(d/\ell)}{8\pi\epsilon ~\ell^4}
{}~\int d^2r\quad [m^z({\bf r})]^2
\label{eq110}
\end{equation}
where $\eta(d/\ell)$ is an exchange correction \cite{macd-platzman-boebinger}.
This acts as an `easy plane' anisotropy converting the system from
Heisenberg to XY symmetry.  For finite $U_z$, $S^2$ is no longer a good
quantum
number. However to a first approximation, the primary effect of $U_z$ is
simply to lift
the degeneracy of the Heisenberg ground states, preferentially selecting
states
with magnetization lying in the XY plane in order to minimize the charging
energy.  The system still has $U(1)$ invariance for rotation about the
$\hat{z}$ isospin axis; {\it i.e.} the number of electrons in each layer is
a good quantum number.
The collective
mode
spectrum is still gapless but the dispersion is now that of a linear
Goldstone
mode \cite{macd-platzman-boebinger,wenandzee,macd-zhang}.
 In the regime of spontaneously broken $U(1)$
symmetry we expect, as discussed below, a finite gap for charged
excitations and
hence a discontinuity $\Delta\mu$
in chemical potential vs. filling factor which implies a
dissipationless quantum Hall state.

If the spacing $d$ exceeds a critical value $d^\ast$ (Fig.~\ref{fig1})
the
system is unable to support a state with strong interlayer correlations and
the
spontaneous $U(1)$ symmetry breaking is destroyed by quantum fluctuations
\cite{macd-platzman-boebinger}.  At this same point we expect the quantum Hall
effect to be
destroyed as the fermionic gap $\Delta\mu$ collapses.
In the Hartree-Fock approximation\cite{harfok} there is a  continuous phase
transition with a gap collapse occurring at finite wavevector so that the
ground
state becomes a charge-density-wave (CDW). Numerical exact diagonalization
calculations suggest  that there is no CDW state but  the
transition {\it is} continuous\cite{ustobepub}.  Little else about
this
transition is firmly established at present.
The most naive picture that the transition is in the universality class of
the
$2 + 1$-D XY model may be incorrect because of the nearby
gapless fermionic degrees of freedom and the fact that (see
below)
vortices carry fermionic charge.

The XY symmetry and finite stiffness in the regime $0<d<d^\ast$ imply a
finite temperature KT transition.  The local XY spin orientation can be
described by a single angle variable $\theta ({\bf r})$ and the effective
Hamiltonian is
\begin{equation}
H=\frac{1}{2}~\rho_s\int d^2r~\vert {\bf\nabla}\theta\vert^2
\label{eq120}
\end{equation}
where, in the Hartree-Fock approximation
$
\rho_s = e^2/(16\pi\epsilon\ell) \int_0^\infty dx\, x^2
\exp (-x^2/2 - dx/\ell).
$
The KT critical temperature (within the Villain approximation)
is \cite{ustobepub}
$T_{\rm
KT}=(\pi /2)~\rho_s\sim 0.5{\rm K}$ for typical sample parameters of Murphy
et al. \cite{murphytobepub}.  A more detailed estimate will require knowledge
of
quantum fluctuation corrections to the stiffness and knowledge of the
vortex
fugacity (core energy) which will be modified by the fact that the vortices
carry electron charge $\pm e/2$.  The vortex charge is readily deduced by
noting
that an electron circling the vortex feels a Berry's phase of $\exp{(i\pi
)}=-1$
induced by the $2\pi$ rotation of its isospin.  Thus it sees the vortex as
half a
flux quantum which for $\sigma_{xy}=\nu~e^2/h$ induces charge $\pm\nu
/2=\pm
1/2$ \cite{kagome}.  One can also see this by noting \cite{ustobepub}
that each vortex is one of four flavors of
`meron' (half a skyrmion) with $\pm$ vorticity at infinity
and a large diameter core
(if $d/\ell$ is small) in which the isospins smoothly rotate either up or down
out
of the XY plane.  The defect has half the topological charge and hence half
the fermionic charge of a skyrmion.

Normally the KT transition would be visible only in the channel which
supports
``isospin supercurrents'', namely {\em oppositely} directed charge currents
in
each layer.  However in the present case a uniform current flow in the same
direction in each layer produces a Hall field which will couple to the vortex
charge.  Thus the KT transition should {\it also}
 manifest itself as a sudden drop in
dissipation in this channel
for $T\leq T_{\rm KT}$.  In contrast to the usual case in
superfluid
and superconducting films however, the linear dissipation will {\em not}
drop to
zero (in this channel).
This is because gapped non-vortex ({\it i.e.\/}, unconfined) fermionic
excitations
will also couple to the Hall field producing weak thermally activated
dissipation.

Another way in which the symmetry can be modified results from the fact that it
is
possible to unbalance the charge density in the two layers by addition of a
bias
field ${\cal E}$ from a gate.  This corresponds to a perturbation
$V_z=e{\cal
E}d~S^z$.  If the symmetry has already been lowered from $SU(2)$ to $U(1)$
by
the finite $d$ this perturbation causes no dramatic changes.  It is
expected
however to renormalize the stiffness $\rho_s$ downwards.  The most naive
picture
of this is simply that when the isospins tilt up in the $\hat{z}$ direction
their
projection onto the XY plane is reduced.  Thus we expect the $\nu =1$ XY
ordered
state to be robust in the presence of variations in $N_\uparrow -
N_\downarrow$.  This is in sharp contrast to the situation at other filling
factors such as $\nu =1/2$ which is described by the Halperin 331 state
\cite{yoshmg} and which requires equal numbers of electrons in each layer.
This
characteristic feature of the $\nu =1$ state appears to have been observed
\cite{murphytobepub}.  We note that the reduction in $\rho_s$ will
renormalize
$T_{\rm KT}$ downwards.  Near $d^\ast$, charge unbalance could be used to tune
to the
quantum critical point without having to vary ($d - d^\ast$).

The final way that one can modify the isospin symmetry is through a finite
tunneling amplitude $t$ which will destroy the XY symmetry by adding
a
term to the Hamiltonian which selects out a preferred direction and thereby
induces an isospin excitation gap
\begin{equation}
H_T=-\int d^2r~{\bf h}\cdot {\bf S}({\bf r})
\label{eq130}
\end{equation}
where ${\bf h}=2t~\hat{x}$.  This perturbation attempts to maximize
$\langle
S^x\rangle$ since it prefers the particles to be in the symmetric tunneling
state.  In the presence of tunneling, the system always has a finite value
for
$\langle S^x\rangle$ no matter how large $d$ is (or even how high the
temperature).  The fermionic gap is stabilized by this effect so that the
critical spacing $d^\ast$ is enhanced
\cite{macd-platzman-boebinger}
in the presence of tunneling
(see Fig.~\ref{fig1}).  However the fermionic gap collapse at the enhanced
$d^\ast$ is
now no longer directly associated with the complete
loss of spontaneous XY magnetic order.

The third axis in Fig.~\ref{fig1} is tilt of the magnetic field achieved by
adding a field $B_\parallel$ in the plane of the layers.  Tilting the field
has
traditionally  been an excellent method of distinguishing
effects of
(real) spins because orbital motion is primarily sensitive to $B_\perp$
while
the (real) spin Zeeman splitting is proportional to the full magnitude of
$B$.   Thus parallel field can destablize spin singlet states in favor of
ferromagnetic states.
For the case of the double layer $\nu =1$ systems studied by Murphy et al
., \cite{murphytobepub} the
ground state is known to already be an isotropic ferromagnetic state  of
the {\it true spins} and the addition of a parallel field will
only serve to further stabilize it.
Nevertheless
these systems are very sensitive to $B_\parallel$.  The activation energy
drops rapidly by up to an order-of-magnitude with increasing $B_\parallel$.
At $B_\parallel = B_\parallel^\ast$ there appears to be
a phase transition to a new state
and the activation gap is independent of further increases in $B_\parallel$.

The effect of $B_\parallel$ on the {\it isospin\/} system can be visualized in
two
different pictures.  We use a gauge in which ${\bf B}_\parallel =
{\bf\nabla}\times {\bf A}_\parallel$ where ${\bf A}_\parallel = B_\parallel
(0,0,x)$.  In this gauge there is no change in the basis orbitals in each
layer
but the tunneling matrix element acquires a position-dependent phase
$t\rightarrow t~e^{iQx}$ where $Q=2\pi /L_\parallel$ and $L_\parallel =
\Phi_0/B_\parallel d$ is the length associated with one flux quantum
$\Phi_0$
between the layers.  This modifies the tunneling Hamiltonian to $H_T=-\int
d^2r~{\bf h}({\bf r})\cdot {\bf S}({\bf r})$ where ${\bf h}({\bf r})$
`tumbles':
${\bf h}({\bf r})=2t~(\cos{Qx},\sin{Qx},0)$.  The effective XY
model now becomes
\begin{equation}
H=\int d^2r~\Biggl\{ \frac{1}{2}~\rho_s\vert {\bf\nabla}\theta\vert^2 -
{t\over 2\pi\ell^2}~\cos{[\theta ({\bf r}) - Qx]}\Biggr\} ,
\label{eq140}
\end{equation}
which is precisely the Pokrovsky-Talapov (P-T) model \cite{bak} and has a
very
rich phase diagram.  For small $Q$ and/or small $\rho_s$ the phase obeys
$\theta
({\bf r})\approx Qx$ but as $B_\parallel$ is increased the local field
tumbles
too rapidly and a continuous
phase transition to an incommensurate state with broken
translation symmetry occurs.  This is because
at large $B_\parallel$ it costs too much
exchange
energy to remain commensurate and the system rapidly gives up the tunneling
energy in order to return to a uniform state ${\bf\nabla}\theta\approx 0$
which
becomes independent of $B_\parallel$.  Using the parameters of the samples
of
Murphy et al. \cite{murphytobepub} we find (for $ d\ll d^*$
within mean-field theory)
a critical field for the transition
$B_\parallel^\ast = B_\perp ~(2 \ell / \pi d) ( 2 t / \pi \rho_s)^{1/2}
\approx 1.6 {\rm T}$ which is within a factor of two
of the observed value\cite{murphytobepub}.
Note that the observed value $B_\parallel^\ast =0.8{\rm T}$
corresponds in
these samples to a large value for $L_\parallel$: $L_\parallel /\ell\sim
20$
indicating that the transition is highly collective in nature.  A second
strong
indication of the collective nature of the excitation gap is the collapse
of
the observed gap at temperatures much smaller than the gap
\cite{murphytobepub}.  We therefore believe that this scenario of a
collective
commensurate-incommensurate transition explains the phase transition seen
in the
experiments.  Numerical exact-diagonalization calculations on small systems
confirm the existence of this
phase transition and show that the fermionic excitation gap
drops to a much smaller value in the incommensurate phase\cite{ustobepub}.

The P-T
 model also undergoes a finite temperature KT phase transition
for $B_\parallel > B_\parallel^\ast$ which restores the translation symmetry
by means of dislocations in the domain wall structure of the incommensurate
phase \cite{bak}.  Thus there are two separate KT transitions in this system,
one for $t=0$, the other for $t\neq 0$ and $B_\parallel > B_\parallel^\ast$.

We now discuss the commensurate-incommensurate phase
transition from the microscopic point of view.
 At $d=0$ the $B_\parallel = 0$ Landau-gauge many-body ground state
wavefunction is a single Slater determinent in which
the single-body states are the symmetric linear combination
of two single-layer states with the same guiding center.
Phase coherence is established by tunneling between
single-layer states with the same guiding center.
For many purposes this state is still a good approximation to
the ground state at finite $d$ since it
optimizes the tunneling energy and has good correlation energy; an
electron in one layer automatically sees an
exchange-correlation hole in the other layer at the
same place.  (It would remain the exact ground state in the absence of
interactions.) From a microscopic point of view it is the good interlayer
correlations of states with phase coherence which leads to the
broken symmetry in the absence of tunneling.
In the Landau gauge, a parallel field causes tunneling to couple
states in the two layers whose momenta differ by $Q$ and whose
guiding centers therefore differ by $\ell^2 Q$.
Thus, for non-interacting electrons the exact ground state in a parallel field
is one in which the exchange-correlation hole is not directly opposite its
electron but rather shifts away by $\ell^2 Q \hat x$
as the {\bf B} field tilts (so that a field line passing through the
electron
also passes through the hole).  This state maintains all of its
tunneling energy but rapidly loses correlation energy as the field tilts.
At large tilt it is better to give up on the tunneling by shifting the two
layers relative to each other to put the correlation hole back next to its
electron.  This shift can be shown to be the change from commensurate to
incommensurate\cite{ustobepub} states disussed above.

Ezawa and Iwazaki\cite{meissner} have recently proposed that there is a
Meissner
effect which can screen out the applied $B_\parallel$ within a finite
`Josephson screening length'.  We find however that this is incorrect.
Consider the analog of eq.(~\ref{eq140}) for a spatially varying $B_\parallel
(x)$
\begin{equation}
H = \int d^2r~ \Biggl\{ \frac{1}{2}~ \rho_s~ \vert\nabla\theta\vert^2 -
\frac{t}{2\pi\ell^2}~ \cos{[\theta ({\bf r}) - A(x)]}\Biggr\}
\label{eq220}
\end{equation}
where
$\partial_x A(x) = (2\pi d/\Phi_0)~ B_\parallel (x)$.  Taking a weak slowly
varying
field $A(x) = A_q~ \cos{qx}$ and making the harmonic approximation, one
readily
finds that the variationally optimal $\theta (x)$ gives an energy density
\begin{equation}
{\cal E}(A_q) = \frac{1}{2}~ \rho_s~ q ^2\vert A_q\vert^2 = \frac{1}{2}~
\rho_s~ \Biggl( \frac{2\pi d}{\Phi_o}\Biggr)^2~ B_\parallel^2
\label{eq230}
\end{equation}
indicating that the system is an ordinary diamagnet with vanishing Meissner
kernel.  The physical reason for this is simply that it is very inexpensive
at
long wavelengths for the phase to twist.  Thus in the `commensurate phase'
the
gauge invariant phase difference across the junction (and hence the
screening
currents) nearly vanish.

The authors are grateful to S.~Q.~Murphy, J.~P.~Eisenstein, and G.~S.~Boebinger
for showing us their data prior to publication, to
S.~Sondhi and S.~Renn  for stimulating discussions, and to J. Carini and
Mats Wallin for
graphics assistance.
This work was supported by NSF
DMR-9113911 and the Aspen Center for Physics.

\begin{figure}
\caption{Schematic zero-temperature phase diagram (with $d/\ell$ increasing
downwards).  The lower surface is
$d^*$ below which $d > d^*$ and
the interlayer correlations are too weak to support a
fermionic gap, $\Delta\mu$.  The upper surface gives $B_\parallel^*$,
the commensurate-incommensurate phase boundary. As $d$ approaches $d^*$,
quantum fluctuations soften the spin stiffness and therefore increase
$B_\parallel^*$.}

\label{fig1}
\end{figure}


\begin{references}
\bibitem{footnote} Throughout this discussion we will ignore true spin and
assume that the real electron spins are fully polarized by the magnetic
field. This simplifying assumption is not necessarily valid at low fields.
\end{references}
\end{document}